\documentclass[12pt]{article}
\usepackage[a4paper, portrait, margin=0.7in]{geometry} 
\usepackage[utf8]{inputenc}
\usepackage{amsmath} 
\usepackage{amsfonts}
\usepackage{amssymb}

\usepackage[T1]{fontenc} 
\usepackage{mathptmx} 

\usepackage{color,soul} 
\usepackage{secdot}

\usepackage[colorlinks,allcolors=blue,bookmarks=false,hypertexnames=true]{hyperref} 
\usepackage{graphicx} 
\usepackage{enumitem} 
\usepackage{hyperref}

\usepackage[style=base,justification=justified]{caption}
\usepackage{titlesec}
\titleformat{\section}
   {\normalfont\Large\bfseries}{\thesection.}{10pt}{}
\titlespacing*\section{0pt}{12pt}{0pt} 

\usepackage[numbers, super, comma, sort&compress]{natbib}

\usepackage{booktabs}
\usepackage{multirow} 
\usepackage{hyphenat}

\newcommand{\mygraphic}[1]{\includegraphics[height=#1]{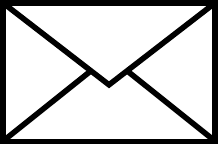}}
\newcommand{\myenv}{\raisebox{0pt}{\mygraphic{.6em}}}
\newcommand{\envelope}[1]{\myenv}

\titleformat*{\subsubsection}{\large\bfseries}

\begin{document}
\sloppy

\noindent{{\LARGE\textbf{\nohyphens{An End-to-End AI-Based Framework for Automated Discovery of CEST/MT MR Fingerprinting Acquisition Protocols and Quantitative Deep Reconstruction (AutoCEST)}}}}

{\large
\noindent \\ Or Perlman\textsuperscript{1*},
Bo Zhu\textsuperscript{1,2*},
Moritz Zaiss\textsuperscript{3,4},
Matthew S. Rosen\textsuperscript{1,2$\#$},
Christian T. Farrar\textsuperscript{1$\#$ $\envelope{}$}
\bigskip
}

\noindent \textsuperscript{1}Athinoula A. Martinos Center for Biomedical Imaging, Department of Radiology, Massachusetts General Hospital and Harvard Medical School, Charlestown, MA, USA

\noindent \textsuperscript{2}Department of Physics, Harvard University, Cambridge, MA, USA

\noindent \textsuperscript{3}Magnetic Resonance Center, Max Planck Institute for Biological Cybernetics, Tübingen, Germany

\noindent \textsuperscript{4}Department of Neuroradiology, Friedrich-Alexander-Universität Erlangen-Nürnberg (FAU), University Hospital Erlangen, Erlangen, Germany

\bigskip

\noindent $^*$O.P. and B.Z. contributed equally to this work.
\newline $^\#$M.S.R. and C.T.F. contributed equally to this work.

\bigskip
\noindent \textbf{$\envelope{}$ Correspondence to: }Christian T. Farrar, Athinoula A. Martinos Center for Biomedical Imaging, Department of Radiology, Massachusetts General Hospital, 149 13th Street, Suite 2301, Charlestown, MA, 02129, USA. email: cfarrar@mgh.harvard.edu

\bigskip
\noindent \textbf{Content: } 
\newline Words in main text: 4685  
\newline Figures: 9 
\newline Tables: 1 
\newline Supporting Information Figures: 1
\newline Supporting Information Tables: 2

\bigskip
\bigskip

\noindent \large{\textbf{Submitted to Magnetic Resonance in Medicine}}

\newpage
{\Large
\noindent\textbf{Abstract}}

\noindent\textbf{Purpose: }To develop an automated machine-learning-based method for the discovery of rapid and quantitative chemical exchange saturation transfer (CEST) MR fingerprinting acquisition and reconstruction protocols.

\noindent\textbf{Methods: }An MR physics governed AI system was trained to generate optimized acquisition schedules and the corresponding quantitative reconstruction neural-network. The system (termed AutoCEST) is composed of a CEST saturation block, a spin dynamics module, and a deep reconstruction network, all differentiable and jointly connected. The method was validated using a variety of chemical exchange phantoms and an in-vivo mouse brain at 9.4T. 

\noindent\textbf{Results: }The acquisition times for AutoCEST optimized schedules ranged from 35-71s, with a quantitative image reconstruction time of only 29 ms. The resulting exchangeable proton concentration maps for the phantoms were in good agreement with the known solute concentrations for AutoCEST sequences (mean absolute error = 2.42 mM; Pearson\textquotesingle s r=0.992, p$<$0.0001), but not for an unoptimized sequence (mean absolute error = 65.19 mM; Pearson\textquotesingle s r=-0.161, p=0.522). Similarly, improved exchange rate agreement was observed between AutoCEST and quantification of exchange using saturation power (QUESP) methods (mean absolute error: 35.8 Hz, Pearson\textquotesingle s r=0.971, p$<$0.0001) compared to an unoptimized schedule and QUESP (mean absolute error = 58.2 Hz; Pearson\textquotesingle s r=0.959, p$<$0.0001). The AutoCEST in-vivo mouse brain semi-solid proton volume-fractions were lower in the cortex (12.21$\pm$1.37\%) compared to the white-matter (19.73$\pm$3.30\%), as expected, and the amide proton volume-fraction and exchange rates agreed with previous reports.

\noindent\textbf{Conclusion: }AutoCEST can automatically generate optimized CEST/MT acquisition protocols that can be rapidly reconstructed into quantitative exchange parameter maps.

\noindent\textbf{Keywords: }Chemical exchange saturation transfer (CEST), magnetization transfer (MT), quantitative imaging, deep learning, magnetic resonance fingerprinting (MRF), optimization.


\newpage
\section{Introduction}
Chemical exchange saturation transfer (CEST) is an increasingly explored molecular imaging technique which allows for the detection of signals associated with milli-molar concentrations of proteins, metabolites, and various molecular compounds \cite{ward2000new, jones2018clinical}. It uses frequency selective radio-frequency (RF) pulses to saturate the magnetization of exchangeable protons on proteins, lipids, and other biologically interesting compounds that later undergo chemical exchange with the bulk water protons, thus altering the MR-detectable signal \cite{van2018magnetization}. 

The potential benefit of using the CEST contrast mechanism was demonstrated in a variety of clinical applications, including cancer detection and grading \cite{zhou2019apt}, stroke characterization \cite{sun2007detection}, characterization of neurodegenerative disorders \cite{bagga2016mapping}, kidney disease monitoring \cite{pavuluri2019noninvasive, longo2021renal}, cartilage and intervertebral disc  imaging \cite{pulickal2019mri, krishnamoorthy2017high}, cell tracking \cite{pumphrey2016advanced, gilad2007artificial, perlman2020redesigned}, and cardiac disease assessment \cite{zhou2016optimized}. 

The most common analysis method for CEST-weighted imaging is the magnetization-transfer-ratio asymmetry (MTR$_{asym}$). Although it can be straight-forwardly calculated and was found useful in many reports, this metric is affected by a mixed contribution from several exchange and relaxation properties, such as the relayed aliphatic nuclear Overhauser enhancement (rNOE) and the water T$_{1}$  relaxation time, that may bias the interpretation of the obtained contrast \cite{zaiss2015relaxation}. Moreover, the MTR$_{asym}$ is strongly affected by the saturation pulse parameters used, challenging the comparison of findings obtained using different protocols, and requiring a rigorous optimization of the acquisition parameters \cite{sun2011simulation}. 

A quantitative CEST technique would clearly be beneficial for overcoming the above-mentioned challenges. The exchange parameters (proton volume fraction and chemical exchange rate) can be quantified by acquiring multiple Z-spectra with different saturation pulse durations and/or powers, followed by analysis using methods such as quantification of exchange using saturation power/time (QUESP/QUEST) \cite{mcmahon2006quantifying}, Omega-plot \cite{dixon2010concentration, wu2015quantitative}, or a full fitting of the Bloch-McConnell equations \cite{zaiss2015combined}. However, the long acquisition times and the complexity of the in-vivo multi-pool environment render this approach sub-optimal for routine clinical use. CEST MR-fingerprinting (MRF \cite{ma2013magnetic}) is a recently suggested promising alternative \cite{cohen2018rapid, zhou2018chemical, heo2019quantifying}. In the MR fingerprinting approach, a pseudo random and fast CEST acquisition schedule is used to obtain different “signal-signatures”, representing different combinations of solute concentration and chemical exchange rate. The acquired experimental signals are then compared to a simulated signal dictionary, allowing the generation of quantitative CEST parameter maps. However, the CEST-MRF performance, and ability to discriminate different exchange rates and proton volume fractions, is critically dependent on the acquisition parameter schedule used \cite{perlman2020cest}. This mandates a careful optimization of the imaging protocol, which is very challenging for CEST/MT imaging given the large number of exchangeable proton pools involved.

The purpose of this work is to develop and validate a novel paradigm for conducting and analyzing CEST experiments. We hypothesized that an MR physics governed AI system, termed here as AutoCEST, can be designed and trained to simultaneously generate an optimized and fast CEST acquisition schedule and at the same time provide the means for reconstructing quantitative exchange-parameter maps, for any given and broadly defined multi-pool CEST/MT scenario. To demonstrate the efficiency and robustness of the method, a validation study using a variety of different CEST phantoms was performed, followed by a proof-of-concept in-vivo mouse imaging experiment.

\section{Methods}
\subsection{AutoCEST architecture and realization}
\begin{figure}[ht!]
\includegraphics[height=6.20in,width=6.92in]{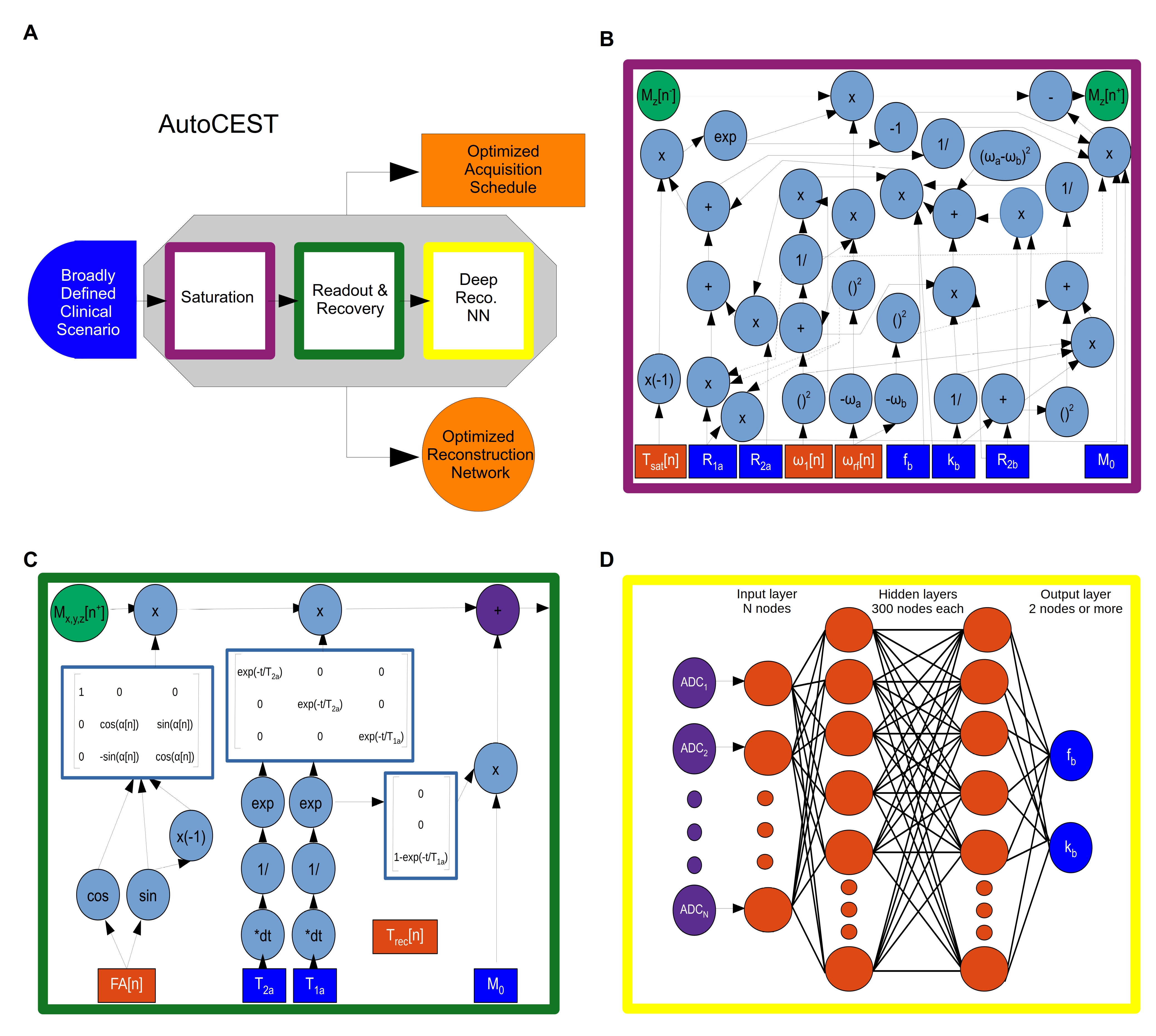}
\captionsetup{aboveskip=0pt,justification=justified}
\caption{(A) Schematic representation of the AutoCEST pre-experiment pipeline. A broadly defined clinical scenario serves as input which allows the experiment optimization by sequentially simulating CEST saturation (purple), readout and recovery (green), and deep reconstruction (yellow). AutoCEST outputs optimized acquisition schedule and an reconstruction network (orange). (B) CEST saturation block as a computational graph. The blue rectangles represent the input intrinsic parameters: initial magnetization (M$_0$), water relaxation rates (R$_{1a}$, R$_{2a}$), solute transverse relaxation (R$_{2b}$), exchange-rate (k$_b$), and volume fraction (f$_b$). The orange rectangles represent the dynamically updated protocol parameters: saturation time (T$_{sat}$), saturation power ($\omega_1$), saturation frequency offset ($\omega_{rf}$). The graph calculates the magnetization at the end of the saturation block M$_z$[n$^+$]. (C) Bloch equation-based image readout as a computational graph. The blue rectangles represent the water-pool parameters, while the orange rectangles represent the dynamically updated protocol parameters: flip angle (FA) and recovery time (T$_{rec}$), which is embedded in the appropriate relaxation step. Note that this is a partial display due to space limitations. (D) Deep reconstruction network for decoding the “ADC” MR signals (purple circles), obtained at C into CEST quantitative parameters (f$_{b}$ and k$_{b}$, blue circles).  } 
\label{Figure1}
\end{figure}

An overview of the AutoCEST approach is described in Figure \ref{Figure1}A. For each chemical exchange scenario of interest (e.g., amide, amine, creatine, magnetization transfer (MT), etc.), the system gets as input a general description of the expected range of parameter values and simulates the expected MR signals from a random CEST acquisition protocol. The system then performs automatic optimization, which ultimately outputs a refined set of acquisition protocol parameters (Figure \ref{Figure1}B,C orange rectangles) as well as optimized neural network weights (Figure \ref{Figure1}D, orange circles), capable of transforming the measured signals into quantitative CEST/MT proton exchange parameter maps. 			

The proposed technique is based on the integration of CEST physics and spin dynamics with deep learning. In a classic neural network, each of the nodes contains a ‘weight element’, which is updated and optimized during the backward propagation step.  To allow an analogous equivalent update of the CEST experiment parameters and achieve efficient optimization using auto-differentiation, the analytical solution of the governing spin dynamics for every step of the imaging experiment was represented as a computational graph (Figure \ref{Figure1}B,C). Next, a deep reconstruction network \cite{cohen2018mr} was used to obtain quantitative CEST/MT parameter maps (proton volume fraction and exchange rate). Notably, the acquisition and reconstruction steps are serially connected to allow joint optimization using automatic differentiation and stochastic gradient descent. The detailed AutoCEST steps include:
\subsubsection{CEST saturation block} The analytical solution of the Bloch-McConnell equations for continous wave RF irradiation,  for either a 2-pool \cite{zaiss2013exchange} (water and solute proton pools) or a 3-pool \cite{zaiss2015combined} (water, solute, and semi-solid/MT proton pools) imaging scenario was represented as a computational graph (Figure \ref{Figure1}B). This allows the calculation of the water-pool M$_z$ component at the end of the saturation, and more importantly, the update of the saturation-block parameters (Figure \ref{Figure1}B, orange rectangles) during training.
\subsubsection{Readout and relaxation spin dynamics module}  In the next step of the forward-direction modeling, the transverse spin components are zeroed-out, assuming sufficient gradient spoiling is applied. Next, the spin dynamics are calculated during excitation and relaxation, using the Bloch equations with a discrete-time state-space model in the rotating frame \cite{zhu2019automated} (Figure \ref{Figure1}C). This allows for the update of the flip-angle (FA) and the recovery time (T$_{rec}$) parameters as well as the calculation of the expected “ADC” signals. 
\subsubsection{Deep reconstruction network} The resulting MR signals are 2-norm normalized along the temporal dimension in a pixel-wise manner and mapped into CEST quantitative parameters using a fully connected 4-layer deep reconstruction network \cite{cohen2018mr} (Figure \ref{Figure1}D). The neural network is composed of a series of fully-connected dense layers, with two hidden layers of 300 nodes each and activated by hyperbolic tangent (tanh) functions.

The entire pipeline was implemented using PyTorch 1.0.1 and Python 3.6.8 on a Linux laptop computer equipped with an 8-core Intel i7-7700HQ CPU (2.80 GHz).  AutoCEST was trained for a variety of chemical exchange scenarios as described in section 2.2, 2.4.2, and Supporting Information Table S1. For each scenario, acquisition schedules of N=10 raw (molecular information encoding) images were generated. The batch size was set to 256 and the number of training epochs set to 100 \cite{zhu2018image}, while a different development set of simulated signals (not included in the training data) was used to confirm that over-fitting is not reached. To further promote robust learning, white Gaussian noise (standard deviation of 0.002) was injected into the training data \cite{zur2009noise, chen2020vivo}. The loss was defined as the mean-squared-error between the estimated proton exchange rate and volume fraction values and their corresponding ground-truth values. The RMSprop algorithm \cite{graves2013generating} was used as the optimizer, with the learning rates of the acquisition schedule parameters and the reconstruction network set to 0.001 and 0.0001, respectively. 
	
To provide basic intuition on the optimization process, AutoCEST was set to update only the saturation pulse power for some of the scenarios (Iohexol, BSA, and in-vivo amide). Next, 2,3, or 5 different acquisition parameters were defined in a simultaneous parameter optimization for the in-vivo MT, pCr, and L-arginine scenarios, respectively.

\begin{figure}[ht!]
\includegraphics[height=3.00in,width=6.92in]{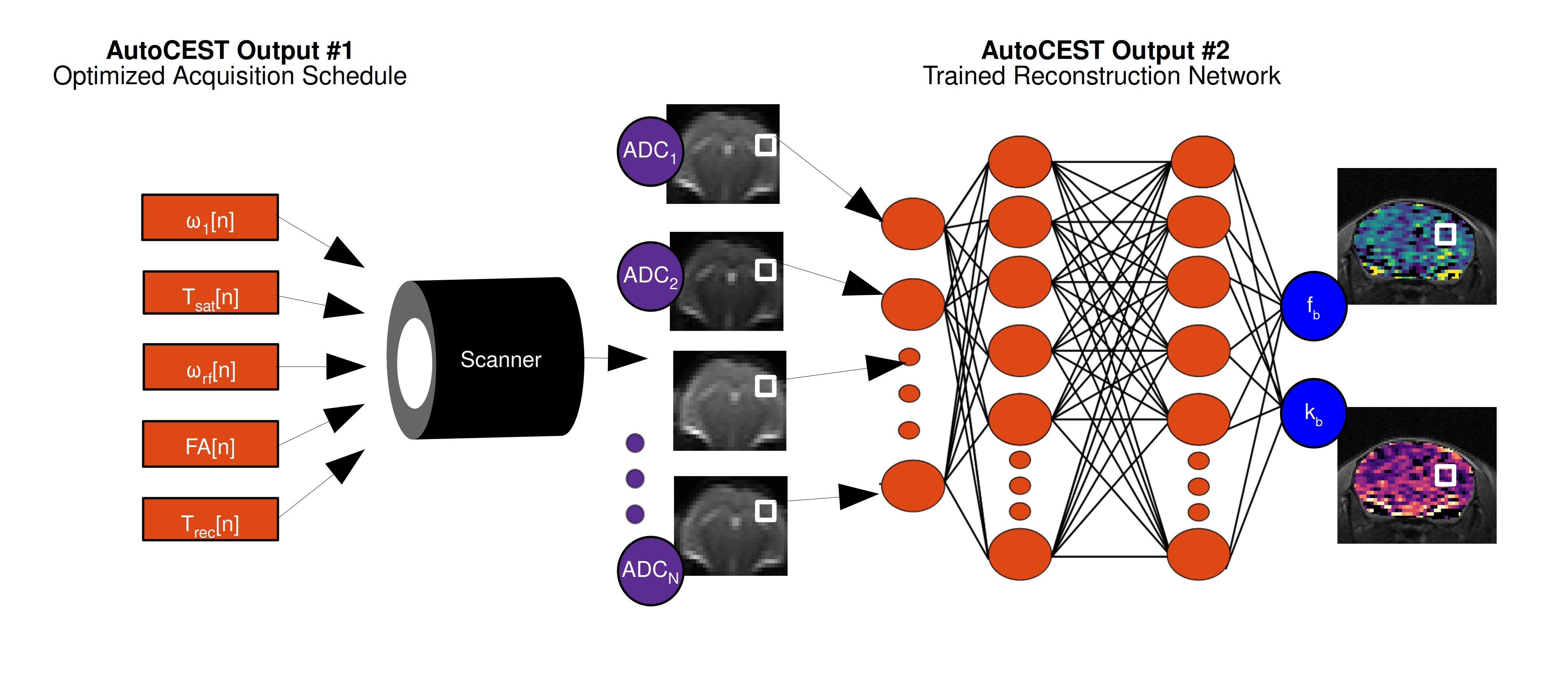}
\caption{AutoCEST-based quantitative image reconstruction. The optimized protocol parameters (orange rectangles, $\omega_{1}$ = saturation pulse power, T$_{sat}$ = saturation pulse duration, $\omega_{rf}$ = saturation pulse frequency offset, FA = readout flip angle, T$_{rec}$= recovery time) are loaded into the scanner, allowing the acquisition of N raw ADC (molecular information encoding) images. The images are fed voxelwise into the trained reconstruction network (orange circles), resulting in quantitative CEST/MT parameter maps (e.g., proton volume fraction f$_b$ and exchange rate k$_b$).}
\label{Figure2}
\end{figure}

Finally, the optimal acquisition schedule parameters found by AutoCEST are loaded into the MR scanner, and a set of N, molecular information encoding, raw images are acquired (Figure \ref{Figure2}). The resulting images are then fed voxel-wise into the AutoCEST trained reconstruction network, resulting in quantitative CEST/MT maps of the imaged subject. 

\subsection{Phantom preparation}
To validate the suggested approach, an extensive in-vitro imaging study was performed using a set of 7 imaging phantoms, each composed of 3 different vials of a particular CEST compound, dissolved in PBS or in a buffer titrated to a particular pH value between 4.0-7.4. The compound concentrations were varied between 12.5-100 mM in all cases except for BSA, where the w/w concentration was varied between 7.5-15\% \cite{banay1992protein, ray2016determination, zollner2018ammonia, xu2014variable}. To verify the AutoCEST robustness for various imaging scenarios, the following compounds were used:
\subsubsection{Iohexol} 
An x-ray iodinated contrast agent, used as a CEST agent for extracellular pH quantification. Iohexol contains two exchangeable amide protons at a chemical shift of 4.3 ppm relative to the resonance frequency of water \cite{longo2016vitro, anemone2017mri}.
\subsubsection{Phosphocreatine (pCr)} A crucial metabolite for heart and skeletal muscle energetics, contains a single guanidinium exchangeable proton at 2.6 ppm \cite{chung2019chemical, chen2020vivo, pavuluri2020amplified}.
\subsubsection{L-arginine} An amino acid with three equivalent exchangeable amine protons with a chemical shift of 3 ppm with respect to the water resonance.
\subsubsection{Bovine serum albumin (BSA)} A protein with a large number of exchangeable amide (3.5 ppm), amine ($\sim$2.75 ppm), and rNOE ($\sim$-3.5 ppm) protons. 

While Iohexol, pCr and L-arginine contain additional exchangeable protons at other chemical shifts than mentioned above, the optimization was focused on their commonly targeted exchangeable protons. To demonstrate the ability of detecting multiple CEST targets within the same phantom, various AutoCEST-based acquisition schedules were generated for imaging the amide, amine, and rNOE exchangeable protons of BSA.

\subsection{Animal preparation} All animal experiments and procedures were performed in accordance with the NIH Guide for the Care and Use of Laboratory Animals and were approved by the Institutional Animal Care and Use Committee of the Massachusetts General Hospital. A C57/BL6 wild-type male mouse (28 gr) was purchased from Jackson Laboratory. It was anesthetized using 1-2\% isoflurane and placed on an MRI cradle with ear and bite bars to secure the head. Respiration rate was monitored with a small animal physiological monitoring system (SA Instruments, Stony Brook, NY), and the temperature was maintained by blowing warm air in the bore of the magnet.

\subsection{Magnetic resonance imaging} All imaging experiments were conducted using a 9.4T MRI scanner (Bruker Biospin, Billerica, MA), employing an in-house programmed, flexible CEST-EPI protocol \cite{cohen2018rapid, perlman2020cest, perlman2020ai}, loaded with the acquisition parameters generated by AutoCEST.  

\subsubsection{Phantom studies}
Imaging was performed using a transmit/receive volume coil (Bruker Biospin, Billerica, MA), a field of view (FOV) of 32 $\times$ 32 mm$^{2}$, a matrix of 64 $\times$ 64 pixels, and a 5 mm slice thickness. The Iohexol and L-arginine phantoms were imaged at room temperature. The pCr and BSA phantoms were heated to 37$^{\circ}$C, using a feedback loop between a small animal physiological monitoring system (SA Instruments, Stony Brook, NY) and a warm air blower. Each phantom was imaged using the AutoCEST-generated scenario-specific acquisition schedules (Figure \ref{Figure3} and Supporting Information Table S1). Single-shot QUESP-EPI images were acquired with saturation at $\pm$1$\times$ the chemical shift of each phantom’s exchangeable proton, except for the BSA where the existence of both the amide and rNOE pools is incompatible with QUESP estimation of the exchange rate. The QUESP saturation pulse powers ranged from 0-6 $\mu$T in 1 $\mu$T increments, the saturation pulse length (T$_{sat}$) was 3s, flip angle (FA) = 90$^{\circ}$, and echo/repetition times (TE/TR) = 20/15000 ms. For comparison, a CEST-MRF scan was performed, using a previously reported phantom acquisition schedule (Supporting Information Figure S1) \cite{cohen2018rapid}, shortened to include only the first N=10 images, for proper comparison with AutoCEST schedules of the same length. The CEST-MRF protocol included a single saturation frequency offset (aimed at the target compound chemical shift frequency), TE/TR = 20/4000 ms, T$_{sat}$ = 3s, and FA = 60$^{\circ}$. A traditional Z-spectra was obtained using a CEST-EPI protocol, employing a saturation pulse power of 2 $\mu$T, T$_{sat}$ = 3s, TE/TR = 20/8000 ms, and saturation frequency offsets of 7 to -7 ppm with 0.25 ppm increments. For calculation of the static magnetic field B$_{0}$ map using the water saturation shift referencing (WASSR) method \cite{kim2009water}, the CEST scan was repeated with a saturation pulse power of 0.3 $\mu$T, and frequency offsets ranging between 1 to -1 ppm with 0.1 ppm increments. T$_{1}$ maps were acquired using the variable repetition-time rapid acquisition with relaxation enhancement (RARE) protocol, with TR = 50, 200, 400, 800, 1500, 3000, 5000, and 7500 ms, TE = 7.2 ms, RARE factor = 2. T$_2$ maps were acquired using the multi-echo spin-echo protocol, TR = 2000 ms, and 25 TE values between 20-500 ms. 

\subsubsection{In vivo study} A quadrature volume coil was used for RF transmission and a mouse brain phased array surface coil was used for receive (Bruker Biospin, Billerica, MA). A field of view (FOV) of 19 $\times$ 19 mm$^{2}$, a matrix of 64 $\times$ 64 pixels, and a 1 mm slice thickness were used in all scans except for a high resolution T$_2$-weighted scan, where the matrix size was set to 128$\times$128, and the TE/TR were 30/2000 ms. MT and amide AutoCEST scans were performed using the generated acquisition schedules described in Figure \ref{Figure3} and Supporting Information Table S1, with an echo time of 21.88 ms.

\subsection{Data analysis}
Raw AutoCEST-generated images were given as input to the trained reconstruction network, yielding the corresponding proton exchange rate and volume fraction maps. T$_{1}$ and T$_{2}$ exponential fitting were performed using a custom‐written program. Conventional CEST images were corrected for B$_{0}$ inhomogeneity using the WASSR method \cite{kim2009water,  liu2010high}. The MTR$_{asym}$ was calculated using: MTR$_{asym}$ = (S$^{- \Delta\omega}$ – S$^{+\Delta\omega}$) / S$_0$, where S$^{\pm\Delta\omega}$ is the signal measured with saturation at $\pm$ the relevant solute chemical shift and S$_0$ is the unsaturated signal.  Exchange rate ground-truth estimation was performed by fitting the QUESP data with the known solute concentration and measured water T$_{1}$  given as fixed inputs for each phantom vial \cite{zaiss2018quesp}. In addition, simultaneous QUESP estimation of both the exchange rate and the unconstrained solute concentration was performed for comparison. 
	
CEST-MRF signal matching was performed by calculating and finding the maximum dot-product (after 2-norm normalization) of each pixel\textquotesingle s trajectory with all relevant simulated dictionary entries. The dictionaries were built using the same data properties used for training AutoCEST (Supporting Information Table S1). Dictionary generation was performed using a numerical solution of the Bloch-McConnell equations, implemented in MATLAB R2018a (The MathWorks, Natick, MA) \cite{cohen2018rapid}. 

In-vitro statistics were calculated using 79 mm$^{2}$ circular regions of interest (ROIs) drawn on each phantom vial. In-vivo statistics were calculated using a gray matter (GM) ROI positioned on the cortex and a white matter (WM) ROI comprised of the corpus callosum and fiber tracts (cerebal peduncle, optic tract, and fimbria) regions. Localization of mouse brain regions was performed using the Allen Mouse Brain Atlas (adult mouse P56, coronal, image 78) as a reference \cite{lein2007genome, AllenMouseIm78}. Pearson\textquotesingle s correlation coefficients were calculated using the open source SciPy scientific computing library for Python \cite{virtanen2020scipy}. Absolute error was defined as |ground truth value $-$ estimated value|. Differences were considered significant at p$<$0.05. 

\section{Results}

\subsection{AutoCEST-generated acquisition protocols}
The AutoCEST optimization of a quantitative acquisition protocol took between 22 min and 5.58 hrs (see Supporting Information Table S1). The optimized protocol acquisition time was 71.1s for pCr, 47.6s for L-arginine, and 35s for all others (Iohexol, BSA amide, BSA amine, BSA rNOE, in-vivo amide, and in-vivo MT). The optimized protocol parameters are shown in Figure \ref{Figure3}. 

\subsection{Phantom study - exchange parameters quantification performance} AutoCEST reconstruction time, for each pair of quantitative proton exchange rate and volume fraction maps (in-vitro and in-vivo) was 28.62 $\pm$ 0.01 ms. The resulting maps for Iohexol, pCr, and L-arg are shown in Figures \ref{Figure4}, \ref{Figure5}, and \ref{Figure6}, respectively. In all cases, an excellent agreement between the AutoCEST-based calculated solute concentrations and the known solute concentrations was obtained, yielding an absolute error of 2.42 ± 2.53 mM and a significant correlation (Pearson\textquotesingle s r=0.992, p$<$0.0001). There was also a significant correlation between the QUESP-calculated and AutoCEST measured proton exchange rates (r=0.971, p$<$0.0001), with an absolute error of 35.8 $\pm$ 29.3 Hz (Supporting Information Table S2). 

The measured solute concentrations obtained with a pseudo-random, unoptimized CEST-MRF acquisition schedule (Figure \ref{Figure4}D,I; Figure \ref{Figure5}D; Figure \ref{Figure6}D,I,N) were poorly correlated with the known solute concentrations (Pearson\textquotesingle s r=-0.161, p=0.522), yielding an absolute error of 65.19 $\pm$ 34.48 mM. The absolute error between the QUESP-calculated and CEST-MRF measured proton exchange rates (Figure \ref{Figure4}E,J; Figure \ref{Figure5}E; Figure \ref{Figure6}E,J,O) was higher than that obtained using AutoCEST (58.2 $\pm$ 56.76 Hz), yet there was a significant correlation between unoptimized CEST-MRF and QUESP measured exchange rates (r=0.959, p$<$0.0001). The implementation of QUESP for simultaneous estimation of the concentration and exchange rate yielded a higher absolute error in solute concentration estimation compared to AutoCEST (11.03 $\pm$ 7.77 mM), and lower absolute error in proton exchange rate estimation (23.94 $\pm$ 29.54 Hz).

To demonstrate the differences between CEST-weighted and AutoCEST output images, conventional MTR$_{asym}$ images (acquired using a fixed saturation pulse power of 2 $\mu$T) for two L-arginine phantoms are provided in Figure \ref{Figure7}. Although the MTR$_{asym}$ image in Figure \ref{Figure7}B provides a clear contrast difference for different L-arg vials, it cannot provide any definite information on the underlying biophysical mechanism; namely, whether a change in the solute concentration or pH is occurring. Moreover, the use of a single pulse saturation power is sub-optimal for imaging scenarios with a wide possible range of proton exchange rates (or pH). This is demonstrated in Figure \ref{Figure7}D, where an L-arginine vial with fast exchanging protons (pH=6) appears to have a negative contrast, due to insufficient saturation. In contrast, a single AutoCEST imaging protocol was capable of correctly quantifying the exchange parameters and uncovering the chemical exchange property responsible for the change in contrast (Figure \ref{Figure6} G,H,I,M).

AutoCEST quantitative images for the amide, rNOE, and amine exchangeable protons of BSA are shown in Figure \ref{Figure8}. The proton volume fraction maps were in good agreement with the ground truth BSA concentration. AutoCEST-based estimation of the exchange rates yielded parameter values (BSA amide $\sim$45 Hz, BSA rNOE $\sim$15 Hz, BSA amine $\sim$783 Hz) in good agreement with previous literature reports \cite{van2003mechanism, chen2018separating, cohen2018rapid, perlman2020ai}. 

\subsection{AutoCEST of in vivo mouse brain
} AutoCEST-generated quantitative semi-solid and amide exchange parameter maps are shown in Figure \ref{Figure9}. The corresponding GM/WM parameter values are shown in Table \ref{Table1}. The semi-solid proton volume fraction map was in good agreement with the Nissl-stained histology tissue section (Figure \ref{Figure9}D), where neuronal cell bodies of GM are preferentially stained. In Particular, an elevated semi-solid volume fraction was observed at the subcortical WM (19.73$\pm$3.30\%) compared to the GM (12.21$\pm$1.37\%), allowing a clear identification of the corpus callosum and white matter fiber tracts. The obtained values were in good agreement with previous literature reports \cite{van2017rapid, herz2021pulseq}. The semi-solid chemical exchange rate was faster in GM (60.81$\pm$9.28 Hz) compared to WM (46.23$\pm$14.70 Hz), in agreement with the literature \cite{stanisz2005t1, herz2021pulseq, heo2019quantifying}. The AutoCEST generated amide proton volume fractions were 0.29$\pm$0.16\% and 0.40$\pm$0.27\% for the GM and WM, respectively, and are in a generally good agreement with a recent mouse study \cite{perlman2020ai}. The amide proton exchange rates were 60.81±9.28 Hz and 73.02±51.11 Hz, for the GM and WM, respectively, which are in the general range of previously reported values \cite{heo2019quantifying, perlman2020ai, liu2013quantitative}, yet higher than the exchange rate measured using water exchange spectroscopy (WEX) in the rat cortex \cite{van2003mechanism}.

\section{Discussion}
Since its establishment more than 20 years ago, CEST MRI has been increasingly investigated as a promising contrast mechanism for studying a variety of disease pathologies. However, while numerous clinical CEST studies have demonstrated its potential \cite{jones2018clinical}, this technique has not yet been adopted in routine clinical practice. The main barriers for clinical translation have been the typically long image acquisition times, the semi-quantitative nature of the proton exchange-weighted image contrast, which depends on a complex overlay of contrasts from different exchangeable proton pools (MT, rNOE, amide, amine), and the inability to separate out contributions to the CEST contrast from chemical exchange rate and proton volume fraction, both of which may be changing with time and disease progression. A quantitative and rapid imaging approach could drastically improve the clinical applicability of CEST, rendering it as an attractive means for gaining new diagnostic insights. 

A CEST MRF approach could help overcome the above challenges and provide quantitative CEST and MT information \cite{cohen2018rapid, zhou2018chemical, heo2019quantifying}. Recently, it was further combined with deep learning architectures, for rapid MT \cite{kang2021unsupervised, kim2020deep} and CEST/MT \cite{perlman2020ai} fingerprinting. However, previous studies have also demonstrated that the ability to discriminate different exchange parameter values depends critically on the choice of acquisition schedule \cite{cohen2018rapid, perlman2020cest}. In particular, the transfer of a CEST-MRF acquisition protocol from one chemical exchange scenario to another is not straight-forward \cite{perlman2020cest}, requiring a through optimization, validation with appropriate tissue-like phantoms, and expert knowledge of the effect of the acquisition protocol properties on the resulting CEST signals. As demonstrated here, naively taking a random CEST-MRF acquisition schedule, which might be useful for a particular CEST agent (Figure \ref{Figure8}N) and applying it for other compounds/applications, could result in very poor performance. This is demonstrated in Figures \ref{Figure4}-\ref{Figure8}, where poor agreement is observed between the exchange parameters determined from an unoptimized CEST-MRF acquisition schedule and the known ground truth values for Iohexol (Figure \ref{Figure4}), phosphocreatine (Figure \ref{Figure5}), L-arginine (Figure \ref{Figure6}), and BSA (Figure \ref{Figure8}) phantoms.
In contrast, here we demonstrate that AutoCEST can adapt and optimize the acquisition schedule for a variety of distinctly different chemical exchange scenarios, accurately mapping the exchange parameters (Figures \ref{Figure4}-\ref{Figure9}, Supporting Information Table S2). In addition, AutoCEST was able to accurately map the solute concentration and chemical exchange rate in a very short time with acquisition times of only 35-71s and an almost instantaneous reconstruction time of 29 ms. This dramatically reduced scan time could greatly assist in incorporating CEST investigations into routine clinical imaging with minimal interference with workflow or time constraints. 

\begin{figure}[ht!]
\centering
\includegraphics[height=6.2in,width=6.92in]{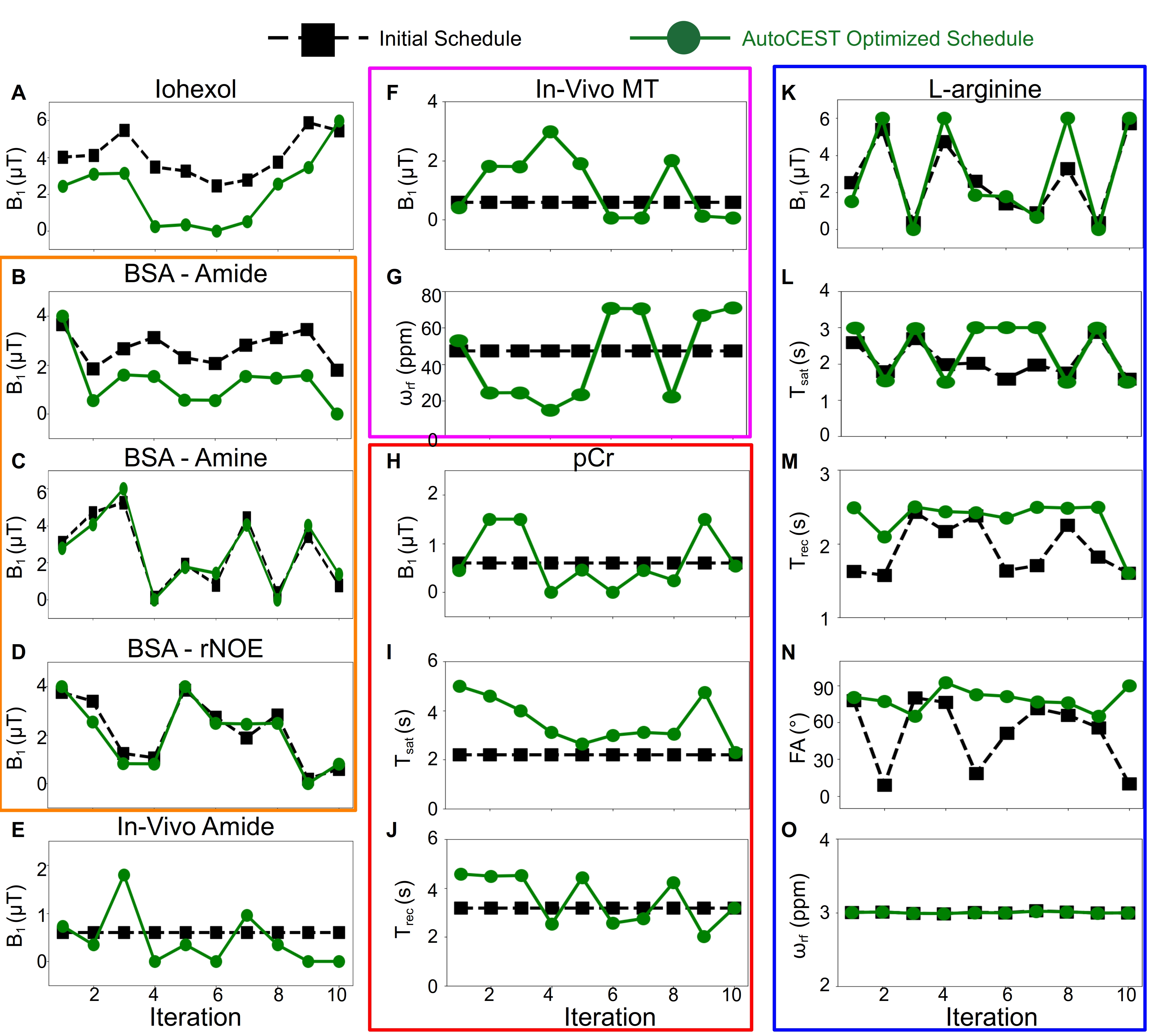}
\caption{AutoCEST-generated acquisition schedules for the various imaging scenarios studied. The black dashed lines with squares represent the random/fixed parameters served for initializing the optimization and the green lines with circles represent the final AutoCEST optimized schedule. For Iohexol (A), BSA - amide (B), BSA – amine (C), BSA – rNOE (D), and In-vivo amide (E), the saturation pulse power (B$_1$) was optimized. For the In-vivo MT (F,G), pCr (H-J), and L-arginine (K-O) cases, 2, 3, and 5 acquisition parameters were simultaneously optimized, respectively. Additional acquisition schedule information is available in Supporting Information Table S1.}
\label{Figure3}
\end{figure}

\begin{figure}[ht!]
\centering
\includegraphics[height=4.0in,width=6.92in]{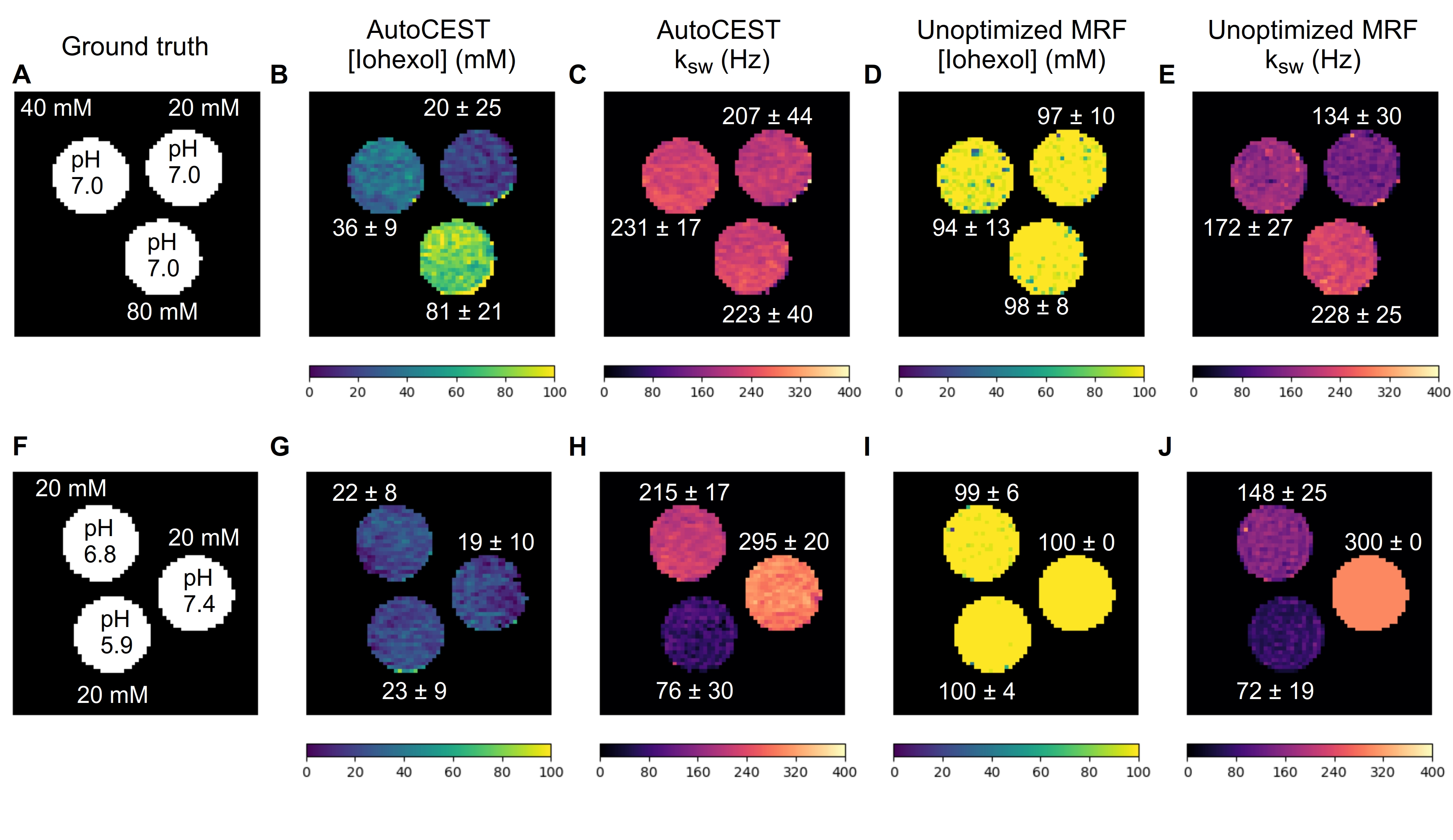}
\caption{Iohexol phantom study. Each row represents a single phantom composed of 3 Iohexol vials, with different concentrations (A) or pH (F). (B, G) AutoCEST-generated Iohexol concentration maps. (C, H) AutoCEST-generated amide (4.3 ppm) proton exchange rate maps. (D, I) CEST-MRF-generated Iohexol concentration maps. (E, J) CEST-MRF-generated amide (4.3 ppm) proton exchange rate maps. The white text next to each vial represent its mean $\pm$ SD parameter value.
}
\label{Figure4}
\end{figure}

\begin{figure}[ht!]
\centering
\includegraphics[height=2.0in,width=6.926in]{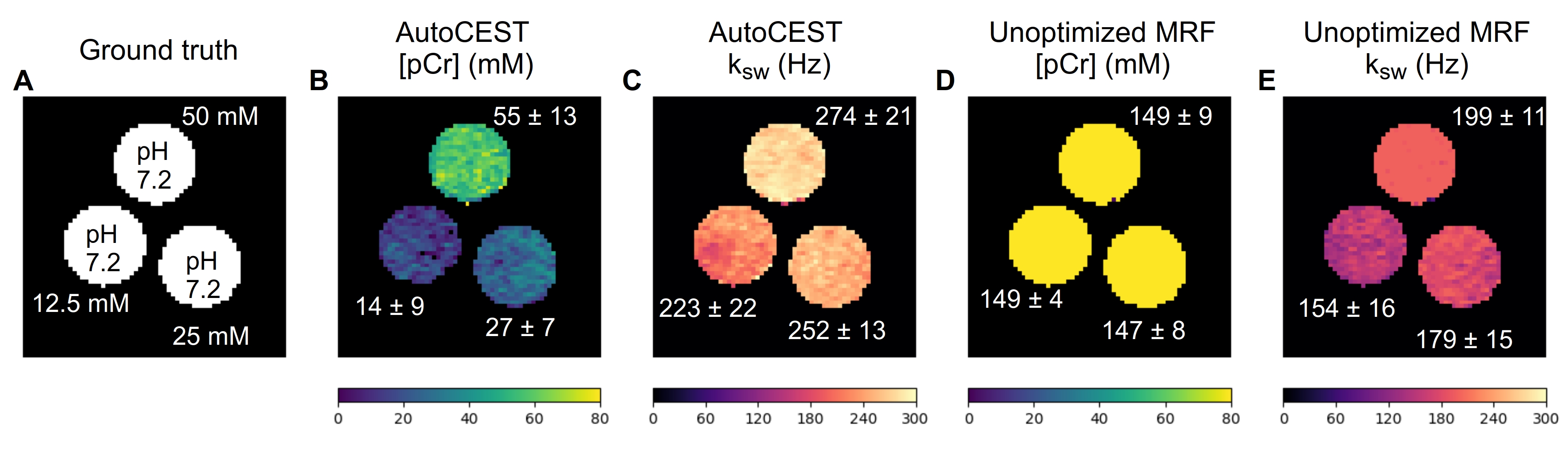}
\caption{Phosphocreatine (pCr) phantom study. (A) Ground truth solute concentration and pH. (B) AutoCEST-generated pCr concentration map. (C) AutoCEST-generated guanidinium (2.6 ppm) proton exchange rate map. (D) CEST-MRF-generated pCr concentration map. (E, J) CEST-MRF-generated guanidinium (2.6 ppm) proton exchange rate map. The white text next to each vial represent its mean $\pm$ SD parameter value.}
\label{Figure5}
\end{figure}

\begin{figure}[ht!]
\centering
\includegraphics[height=6.0in,width=6.92in]{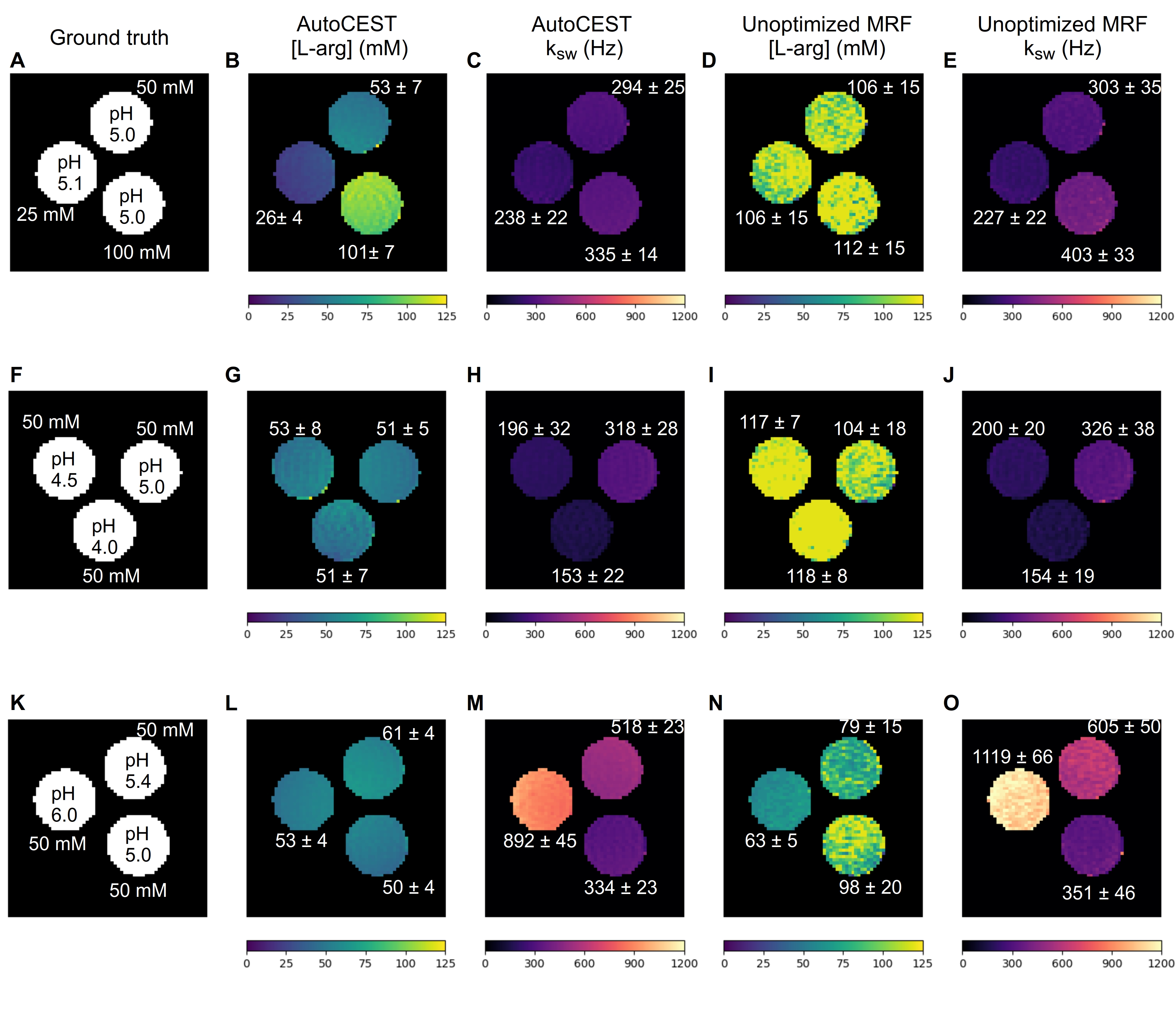}
\caption{L-arginine phantom study. Each row represents a single phantom composed of 3 L-arginine vials, with different concentrations (A) or pH (F, K). (B, G, L) AutoCEST-generated L-arginine concentration maps. (C, H, M) AutoCEST-generated amine (3 ppm) proton exchange rate maps. (D, I, N) CEST-MRF-generated L-arginine concentration maps. (E, J, O) CEST-MRF-generated L-arginine (3 ppm) proton exchange rate maps. The white text next to each vial represent its mean $\pm$ SD parameter value.}
\label{Figure6}
\end{figure}

\begin{figure}[ht!]
\centering
\includegraphics[height=3.80in,width=3.42in]{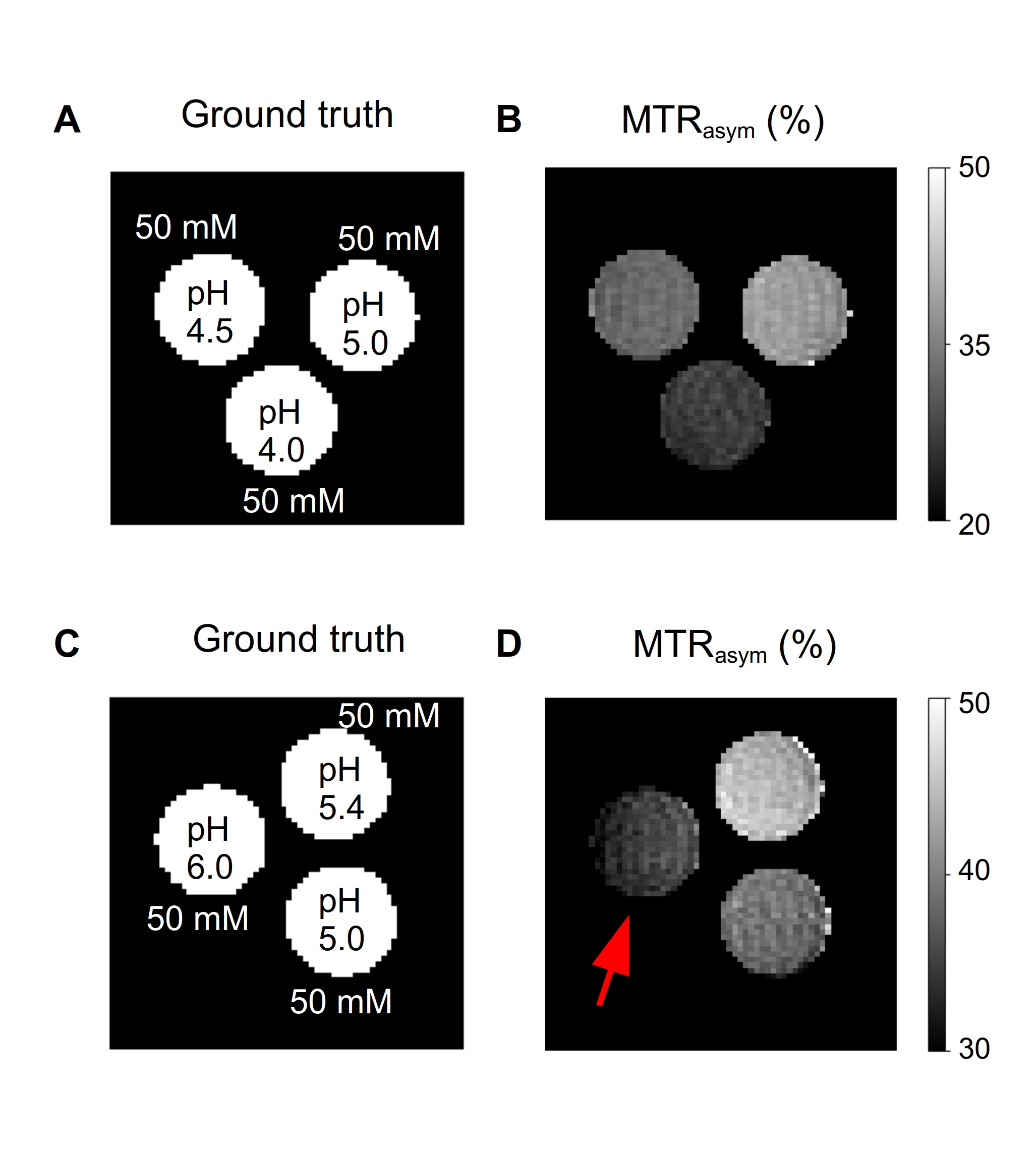}
\caption{Conventional CEST-weighted imaging. Each row represents a single phantom, composed of 3 L-arginine vials, with different pH. (B, D) MTR$_{asym}$ images obtained after Z-spectrum acquisition with a fixed saturation pulse power of 2 $\mu$T. The red arrow in D points to the highest pH vial, which demonstrated negative MTR$_{asym}$ contrast due to insufficient saturation. AutoCEST generated maps of the same phantoms are available in Figure \ref{Figure6}G,H,L,M.}
\label{Figure7}
\end{figure}

\begin{figure}[ht!]
\centering
\includegraphics[height=6.0in,width=6.92in]{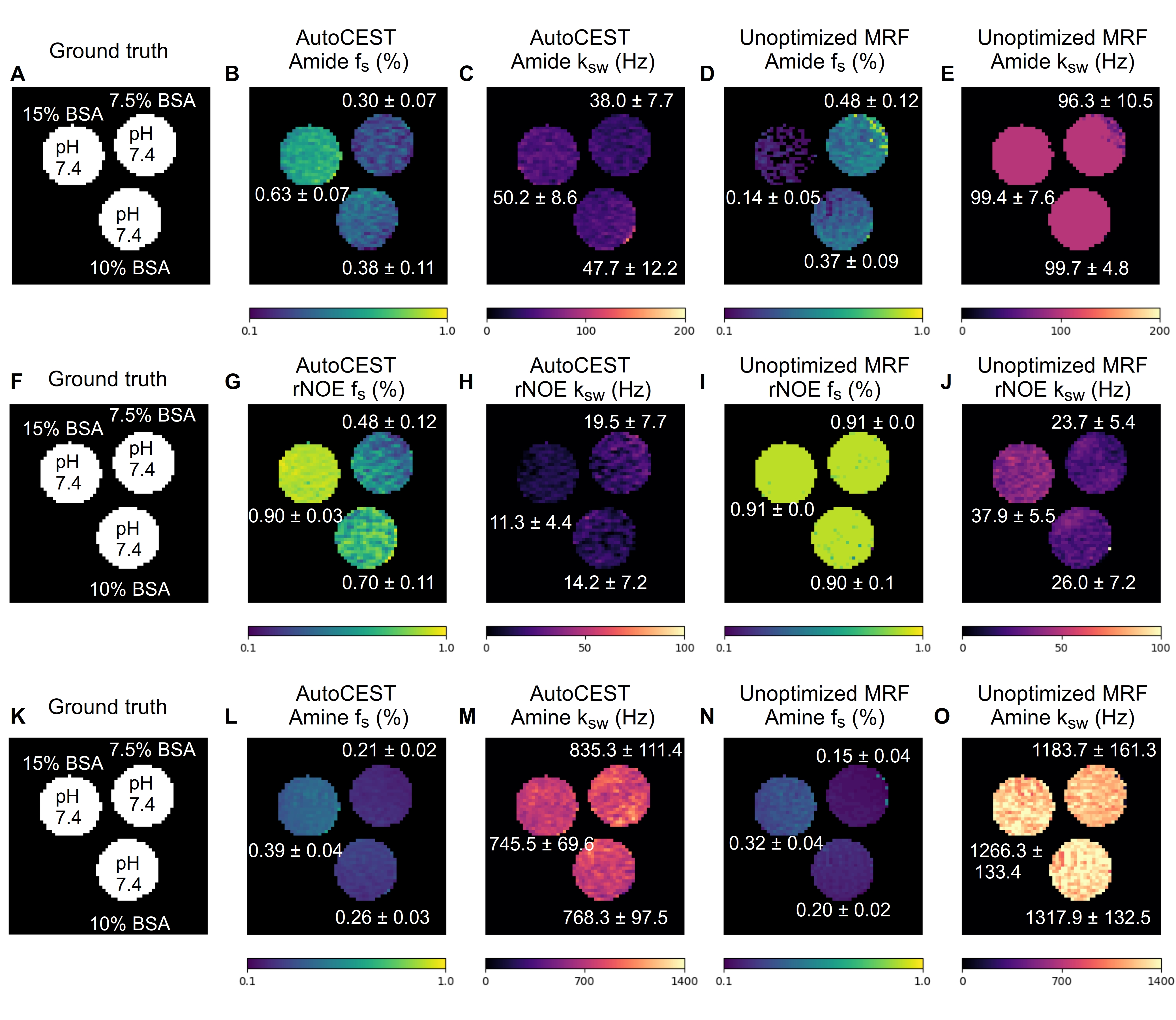}
\caption{BSA phantom study. Each row represents a different molecular target (amide at 3.5 ppm, rNOE at -3.5 ppm, or amine at 2.75 ppm exchangeable protons, respectively), imaged from the same phantom (A, F, K). (B, G, L) AutoCEST-generated amide, rNOE, and amine proton volume fraction maps, respectively. (C, H, M) AutoCEST-generated amide, rNOE, and amine proton exchange rate maps, respectively. (D, I, N) CEST-MRF-generated amide, rNOE, and amine proton volume fraction maps, respectively. (E, J, O) CEST-MRF-generated amide, rNOE, and amine proton exchange rate maps, respectively. The white text next to each vial represent its mean $\pm$ SD parameter value.}
\label{Figure8}
\end{figure}

\begin{figure}[ht!]
\centering
\includegraphics[height=4.0in,width=6.92in]{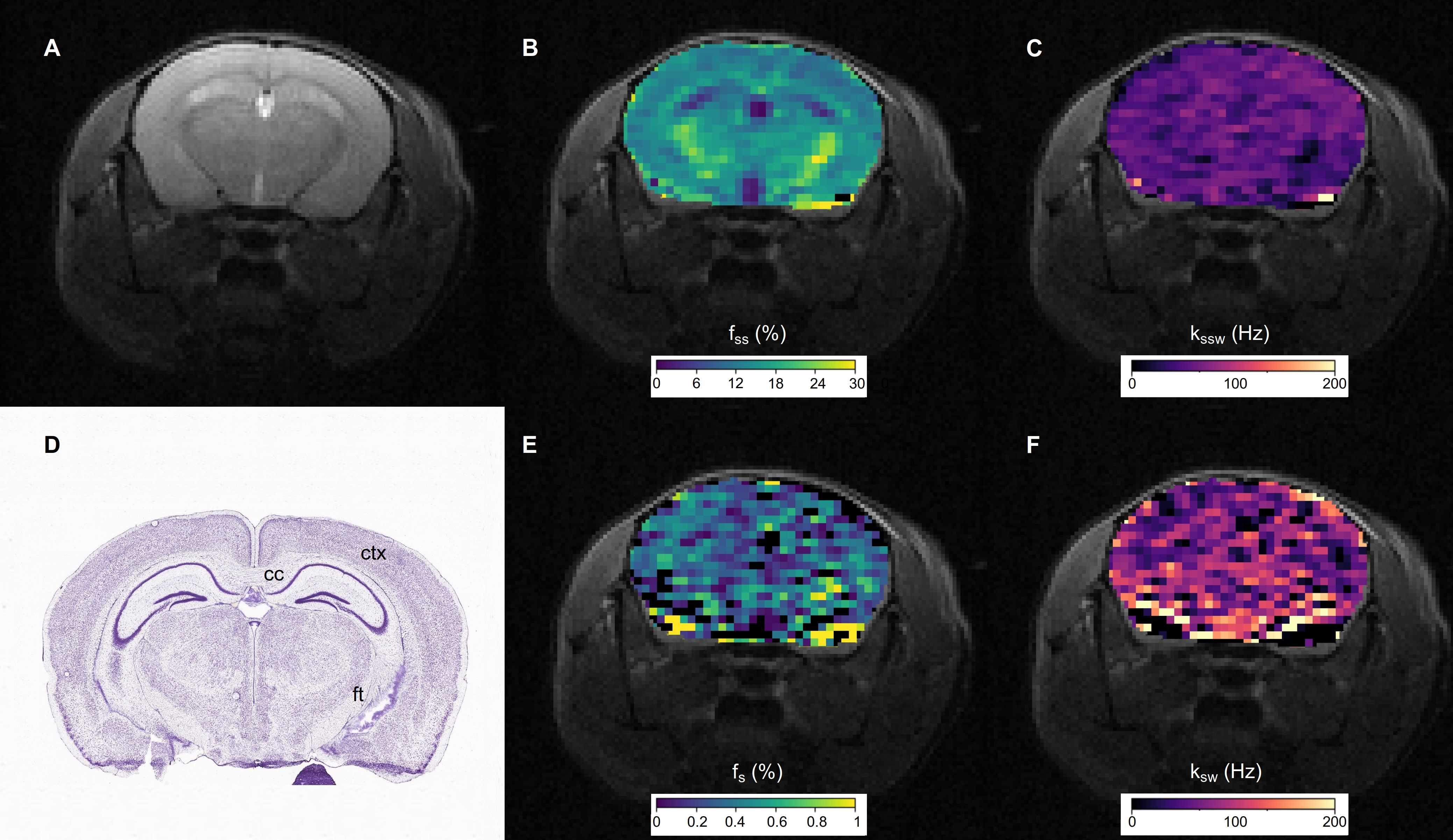}
\caption{AutoCEST imaging of an in-vivo mouse brain. (A) T$_2$-weighted image and (D) corresponding Nissl-stained mouse brain section with the cerebral cortex (ctx), corpus callosum (cc), and fiber tracts (ft, composed of cerebal peduncle, optic tract, and fimbria) identified \cite{lein2007genome, AllenMouseIm78}. AutoCEST-generated (B) semi-solid proton volume fraction and (C) chemical exchange rate maps and amide proton volume fraction (E) and chemical exchange rate maps (F).}
\label{Figure9}
\end{figure}

The AutoCEST method proposed here, constitutes a unified framework for both the design of fast CEST/MT acquisition protocols and the reconstruction of quantitative parameter maps. Importantly, the method is fully automatic, removing the need for user-dependent analysis and exhaustive tuning and optimization of acquisition protocols. The AutoCEST realization was inspired and driven by the AutoSeq method, which allows for automatic sequence generation in 1D and single pixel T$_1$/T$_2$ quantitative imaging \cite{zhu2018automated, zhu2019automated}. Recently, the MRzero \cite{loktyushin2021mrzero} method was reported, which furthers incorporates gradient and RF-events for learning 2D imaging acquisition schedules, including free k-space trajectories \cite{glang2021advances}. The present work expands on the idea of AI-based sequence design for CEST/MT quantitative imaging, where a crucial need for automatic schedule invention lies.
	Observing the differences between the acquisition schedules used for AutoCEST initialization and the final optimized schedules (Figure \ref{Figure3}), can provide some intuition on the underlying optimization performed. For example, optimization of the acquisition schedules for both the Iohexol (at room temperature) and BSA-amide imaging scenarios resulted in saturation pulse powers that were lower than initialized. This can be explained by the relatively slow exchange rates of these compounds (<300 Hz) which are not expected to benefit from a high saturation power. Similarly, the optimal saturation frequency offset for amide and amine exchangeable protons remained roughly fixed at the solute frequency offset, as expected for a CEST agent with a relatively narrow spectral width (Figure \ref{Figure3}O), while the spectrally very broad semi-solid MT case required a wider range of saturation pulse frequency offsets (Figure \ref{Figure3}G).
	
The particular patterns obtained for some of the optimized parameters appeared to lack any noticeable human-intuition (Figure \ref{Figure3}E,F,G,H,I,J), similar to the results obtained in T$_1$/T$_2$ MRF sequence generation \cite{cohen2017algorithm}. This highlights the need for an automated computer-based optimization process. In addition, although the resulting optimized protocols were mostly substantially different than the initial acquisition schedules, there were a few cases where the protocols were not drastically modified (Figure \ref{Figure3}C,D). This might explain the success of some previously reported random CEST-MRF schedules, which could in some cases randomly “land on” suitable parameters. 

The AutoCEST-generated schedules tended to have a longer recovery time compared to the initial value. Notably, quantitative CEST is characterized by an internal trade-off between a sufficiently high SNR and a clinically relevant scan time \cite{perlman2020cest}. While longer recovery times improve the former, some compromise must be made to accommodate for the latter. In this work, we have either fixed or limited the lower and upper bounds for the AutoCEST optimized T$_{sat}$ and T$_{rec}$ (Supporting Information Table S1). Although probably not reaching to the optimal possible sensitivity, this approach has yielded very good performance (Supporting Information Table S2), while satisfying the need for a short scan time with all output schedules shorter than 72s.
 
All the experiments conducted in this work were fixed to create acquisition schedules of N=10 raw images, together with additional restrictions on the scan time (in the form of maximal T$_{rec}$ and T$_{sat}$, Supporting Information Table S1). While this was done to push the boundaries of quantitative CEST beyond the limits set by previous work, a slight relaxation in the parameter restriction could improve the quantitation performance, and still retain sufficiently clinically relevant scan times. In the future, the number of raw images acquired (N) could be defined as a dynamically optimized parameter. In addition, while the saturation power was limited to not exceed a fixed value for each of the scenarios (Supporting Information Table S1), it could be replaced in the future by a specific absorption rate (SAR) penalty term, incorporated in the cost-function \cite{loktyushin2021mrzero}. Similarly, a penalty term for exceedingly long scan times could be used to further improve SNR/scan-time balance.

The AutoCEST determined exchange parameters for the in vivo mouse brain were in general agreement with the literature for 2-pool MT and 3-pool amide/MT imaging; however, the resulting in vivo amide exchange rates were higher than a previous WEX estimation in the rat cortex \cite{van2003mechanism}. Although amide chemical exchange rate is a subject of some controversy in the field, given that various groups have reported amide proton exchange rates $>$100 Hz \cite{heo2019quantifying, wang2020novel}, it might be useful to pursue additional strategies for exploring multi-pool AutoCEST imaging. In particular, the use of a single acquisition schedule, with saturation at the amide proton frequency only, may make discrimination of both amide and MT pool exchange parameters more challenging. For example, we have recently demonstrated that nailing down the MT pool parameters, with an MT specific acquisition schedule, and then sequentially using them as direct inputs for the amide-pool classification, significantly improved the performance in CEST-MRF of oncolytic virotherapy treated mice \cite{perlman2020ai}. Future work could expand the architecture of AutoCEST to allow for such sequentially acquired information to be incorporated.

While the experiments described here were all performed on preclinical scanners with continuous wave saturation pulses, the implementation of AutoCEST for clinical scanners could be straightforwardly translated for cases in which a single continuous-wave block pulse could be applied (e.g., when the required T$_{sat}$ and/or B$_{1max}$ are not expected to be too large), or by modifying the analytical solution of the CEST saturation block to accommodate for a pulse train \cite{meissner2015quantitative, roeloffs2015towards, gochberg2018towards}. 

\section{Conclusion}
The suggested framework provides a fast and automatic means for designing and analyzing quantitative CEST experiments, potentially contributing to the efforts to disseminate CEST/MT in the clinic. The superiority of AutoCEST performance compared to unoptimized CEST MRF highlights the importance of optimizing the acquisition schedule for improved discrimination of the exchange parameters.

\section*{Acknowledgment}
The work was supported by the US National Institutes of Health Grants R01-CA203873 and P41-RR14075. The research was supported by a CERN openlab cloud computing grant. This project has received funding from the European Union\textquotesingle s Horizon 2020 research and innovation programme under the Marie Skłodowska-Curie grant agreement No 836752 (OncoViroMRI). This paper reflects only the author\textquotesingle s view and the Research Executive Agency of the European Commission is not responsible for any use that may be made of the information it contains.

\section*{Data availability statement}
The raw and analyzed AutoCEST data used in this work are available in \href{https://doi.org/10.6084/m9.figshare.14877765}{https://doi.org/10.6084/m9.figshare.14877765}. MR-fingerprinting dictionaries can be reproduced using the open-source code available in \href{https://pulseq-cest.github.io}{https://pulseq-cest.github.io} \cite{herz2021pulseq} with the parameters described in Supporting Information Table S1. Conventional CEST analysis can be performed using the code available in \href{https://github.com/cest-sources}{https://github.com/cest-sources}. Source code is available from the corresponding author upon request.

\nocite{zaiss2011quantitative}
\nocite{zhang2017accuracy}

\newpage
\bibliography{Bibliography} 

\begin{thebibliography}{10}

\bibitem{ward2000new}
Ward KM, Aletras AH, Balaban Robert~S. A new class of contrast agents for MRI
  based on proton chemical exchange dependent saturation transfer (CEST)  {\it
  Journal of magnetic resonance. } 2000;143:79--87.

\bibitem{jones2018clinical}
Jones Kyle~M, Pollard Alyssa~C, Pagel Mark~D. Clinical applications of chemical
  exchange saturation transfer (CEST) MRI  {\it Journal of Magnetic Resonance
  Imaging. } 2018;47:11--27.

\bibitem{van2018magnetization}
Zijl Peter~CM, Lam Wilfred~W, Xu~Jiadi, Knutsson Linda, Stanisz Greg~J.
  Magnetization transfer contrast and chemical exchange saturation transfer
  MRI. Features and analysis of the field-dependent saturation spectrum  {\it
  Neuroimage. } 2018;168:222--241.

\bibitem{zhou2019apt}
Zhou Jinyuan, Heo Hye-Young, Knutsson Linda, Zijl Peter~CM, Jiang Shanshan.
  APT-weighted MRI: Techniques, current neuro applications, and challenging
  issues  {\it Journal of Magnetic Resonance Imaging. } 2019;50:347--364.

\bibitem{sun2007detection}
Sun Phillip~Zhe, Zhou Jinyuan, Sun Weiyun, Huang Judy, Van~Zijl Peter~CM.
  Detection of the ischemic penumbra using pH-weighted MRI  {\it Journal of
  Cerebral Blood Flow \& Metabolism. } 2007;27:1129--1136.

\bibitem{bagga2016mapping}
Bagga Puneet, Crescenzi Rachelle, Krishnamoorthy Guruprasad, et al. Mapping the
  alterations in glutamate with Glu CEST MRI in a mouse model of dopamine
  deficiency  {\it Journal of neurochemistry. } 2016;139:432--439.

\bibitem{pavuluri2019noninvasive}
Pavuluri KowsalyaDevi, Manoli Irini, Pass Alexandra, et al. Noninvasive
  monitoring of chronic kidney disease using pH and perfusion imaging  {\it
  Science advances. } 2019;5:eaaw8357.

\bibitem{longo2021renal}
Longo Dario~Livio, Irrera Pietro, Consolino Lorena, Sun Phillip~Zhe, McMahon
  Michael~T. Renal pH Imaging Using Chemical Exchange Saturation Transfer
  (CEST) MRI: Basic Concept  {\it Preclinical MRI of the Kidney. } 2021:241.

\bibitem{pulickal2019mri}
Pulickal Tina, Boos Johannes, Konieczny Markus, et al. MRI identifies
  biochemical alterations of intervertebral discs in patients with low back
  pain and radiculopathy  {\it European radiology. } 2019;29:6443--6446.

\bibitem{krishnamoorthy2017high}
Krishnamoorthy Guruprasad, Nanga Ravi Prakash~Reddy, Bagga Puneet, Hariharan
  Hari, Reddy Ravinder. High quality three-dimensional gagCEST imaging of in
  vivo human knee cartilage at 7 Tesla  {\it Magnetic resonance in medicine. }
  2017;77:1866--1873.

\bibitem{pumphrey2016advanced}
Pumphrey Ashley, Yang Zhengshi, Ye~Shaojing, et al. Advanced cardiac chemical
  exchange saturation transfer (cardioCEST) MRI for in vivo cell tracking and
  metabolic imaging  {\it NMR in Biomedicine. } 2016;29:74--83.

\bibitem{gilad2007artificial}
Gilad Assaf~A, McMahon Michael~T, Walczak Piotr, et al. Artificial reporter
  gene providing MRI contrast based on proton exchange  {\it Nature
  biotechnology. } 2007;25:217--219.

\bibitem{perlman2020redesigned}
Perlman Or, Ito Hirotaka, Gilad Assaf~A, et al. Redesigned reporter gene for
  improved proton exchange-based molecular MRI contrast  {\it Scientific
  reports. } 2020;10:1--9.

\bibitem{zhou2016optimized}
Zhou Zhengwei, Chen Yuhua, Xie Yibin, et al. Optimized cardiac CEST MRI for
  assessment of metabolic activity in the heart  {\it Journal of Cardiovascular
  Magnetic Resonance. } 2016;18:1--3.

\bibitem{zaiss2015relaxation}
Zaiss Moritz, Windschuh Johannes, Paech Daniel, et al. Relaxation-compensated
  CEST-MRI of the human brain at 7 T: Unbiased insight into NOE and amide
  signal changes in human glioblastoma  {\it Neuroimage. } 2015;112:180--188.

\bibitem{sun2011simulation}
Sun Phillip~Zhe, Wang Enfeng, Cheung Jerry~S, Zhang Xiaoan, Benner Thomas,
  Sorensen A~Gregory. Simulation and optimization of pulsed radio frequency
  irradiation scheme for chemical exchange saturation transfer (CEST)
  MRI—demonstration of pH-weighted pulsed-amide proton CEST MRI in an animal
  model of acute cerebral ischemia  {\it Magnetic resonance in medicine. }
  2011;66:1042--1048.

\bibitem{mcmahon2006quantifying}
McMahon Michael~T, Gilad Assaf~A, Zhou Jinyuan, Sun Phillip~Z, Bulte Jeff~WM,
  Van~Zijl Peter~CM. Quantifying exchange rates in chemical exchange saturation
  transfer agents using the saturation time and saturation power dependencies
  of the magnetization transfer effect on the magnetic resonance imaging signal
  (QUEST and QUESP): pH calibration for poly-L-lysine and a starburst dendrimer
   {\it Magnetic Resonance in Medicine: An Official Journal of the
  International Society for Magnetic Resonance in Medicine. } 2006;55:836--847.

\bibitem{dixon2010concentration}
Dixon W~Thomas, Ren Jimin, Lubag Angelo~JM, et al. A concentration-independent
  method to measure exchange rates in PARACEST agents  {\it Magnetic resonance
  in medicine. } 2010;63:625--632.

\bibitem{wu2015quantitative}
Wu~Renhua, Xiao Gang, Zhou Iris~Yuwen, Ran Chongzhao, Sun Phillip~Zhe.
  Quantitative chemical exchange saturation transfer (qCEST) MRI--omega plot
  analysis of RF-spillover-corrected inverse CEST ratio asymmetry for
  simultaneous determination of labile proton ratio and exchange rate  {\it NMR
  in biomedicine. } 2015;28:376--383.

\bibitem{zaiss2015combined}
Zaiss Moritz, Zu~Zhongliang, Xu~Junzhong, et al. A combined analytical solution
  for chemical exchange saturation transfer and semi-solid magnetization
  transfer  {\it NMR in Biomedicine. } 2015;28:217--230.

\bibitem{ma2013magnetic}
Ma~Dan, Gulani Vikas, Seiberlich Nicole, et al. Magnetic resonance
  fingerprinting  {\it Nature. } 2013;495:187--192.

\bibitem{cohen2018rapid}
Cohen Ouri, Huang Shuning, McMahon Michael~T, Rosen Matthew~S, Farrar
  Christian~T. Rapid and quantitative chemical exchange saturation transfer
  (CEST) imaging with magnetic resonance fingerprinting (MRF)  {\it Magnetic
  resonance in medicine. } 2018;80:2449--2463.

\bibitem{zhou2018chemical}
Zhou Zhengwei, Han Pei, Zhou Bill, et al. Chemical exchange saturation transfer
  fingerprinting for exchange rate quantification  {\it Magnetic resonance in
  medicine. } 2018;80:1352--1363.

\bibitem{heo2019quantifying}
Heo Hye-Young, Han Zheng, Jiang Shanshan, Sch{\"a}r Michael, Zijl Peter~CM,
  Zhou Jinyuan. Quantifying amide proton exchange rate and concentration in
  chemical exchange saturation transfer imaging of the human brain  {\it
  Neuroimage. } 2019;189:202--213.

\bibitem{perlman2020cest}
Perlman Or, Herz Kai, Zaiss Moritz, Cohen Ouri, Rosen Matthew~S, Farrar
  Christian~T. CEST MR-Fingerprinting: Practical considerations and insights
  for acquisition schedule design and improved reconstruction  {\it Magnetic
  resonance in medicine. } 2020;83:462--478.

\bibitem{cohen2018mr}
Cohen Ouri, Zhu Bo, Rosen Matthew~S. MR fingerprinting deep reconstruction
  network (DRONE)  {\it Magnetic resonance in medicine. } 2018;80:885--894.

\bibitem{zaiss2013exchange}
Zaiss Moritz, Bachert Peter. Exchange-dependent relaxation in the rotating
  frame for slow and intermediate exchange--modeling off-resonant spin-lock and
  chemical exchange saturation transfer  {\it NMR in Biomedicine. }
  2013;26:507--518.

\bibitem{zhu2019automated}
Zhu Bo, Liu J, Koonjoo Neha, Rosen B, Rosen Matthew~S. AUTOmated pulse SEQuence
  generation (AUTOSEQ) and neural network decoding for fast quantitative MR
  parameter measurement using continuous and simultaneous RF transmit and
  receive  in {\it ISMRM Annual Meeting \& Exhibition};1090 2019.

\bibitem{zhu2018image}
Zhu Bo, Liu Jeremiah~Z, Cauley Stephen~F, Rosen Bruce~R, Rosen Matthew~S. Image
  reconstruction by domain-transform manifold learning  {\it Nature. }
  2018;555:487--492.

\bibitem{zur2009noise}
Zur Richard~M, Jiang Yulei, Pesce Lorenzo~L, Drukker Karen. Noise injection for
  training artificial neural networks: A comparison with weight decay and early
  stopping  {\it Medical physics. } 2009;36:4810--4818.

\bibitem{chen2020vivo}
Chen Lin, Sch{\"a}r Michael, Chan Kannie~WY, et al. In vivo imaging of
  phosphocreatine with artificial neural networks  {\it Nature communications.
  } 2020;11:1--10.

\bibitem{graves2013generating}
Graves Alex. Generating sequences with recurrent neural networks  {\it arXiv
  preprint arXiv:1308.0850. } 2013.

\bibitem{banay1992protein}
Banay-Schwartz M, Kenessey A, DeGuzman T, Lajtha A, Palkovits M. Protein
  content of various regions of rat brain and adult and aging human brain  {\it
  Age. } 1992;15:51--54.

\bibitem{ray2016determination}
Ray Kevin~J, Larkin James~R, Tee Yee~K, et al. Determination of an optimally
  sensitive and specific chemical exchange saturation transfer MRI
  quantification metric in relevant biological phantoms  {\it NMR in
  Biomedicine. } 2016;29:1624--1633.

\bibitem{zollner2018ammonia}
Z{\"o}llner Helge~J{\"o}rn, Butz Markus, Kircheis Gerald, et al.
  Ammonia-weighted imaging by chemical exchange saturation transfer MRI at 3 T
  {\it NMR in Biomedicine. } 2018;31:e3947.

\bibitem{xu2014variable}
Xu~Jiadi, Yadav Nirbhay~N, Bar-Shir Amnon, et al. Variable delay multi-pulse
  train for fast chemical exchange saturation transfer and relayed-nuclear
  overhauser enhancement MRI  {\it Magnetic resonance in medicine. }
  2014;71:1798--1812.

\bibitem{longo2016vitro}
Longo Dario~Livio, Michelotti Filippo, Consolino Lorena, et al. In vitro and in
  vivo assessment of nonionic iodinated radiographic molecules as chemical
  exchange saturation transfer magnetic resonance imaging tumor perfusion
  agents  {\it Investigative radiology. } 2016;51:155--162.

\bibitem{anemone2017mri}
Anemone Annasofia, Consolino Lorena, Longo Dario~Livio. MRI-CEST assessment of
  tumour perfusion using X-ray iodinated agents: comparison with a conventional
  Gd-based agent  {\it European radiology. } 2017;27:2170--2179.

\bibitem{chung2019chemical}
Chung Julius~Juhyun, Jin Tao, Lee Jung~Hee, Kim Seong-Gi. Chemical exchange
  saturation transfer imaging of phosphocreatine in the muscle  {\it Magnetic
  resonance in medicine. } 2019;81:3476--3487.

\bibitem{pavuluri2020amplified}
Pavuluri KowsalyaDevi, Rosenberg Jens~T, Helsper Shannon, Bo~Shaowei, McMahon
  Michael~T. Amplified detection of phosphocreatine and creatine after
  supplementation using CEST MRI at high and ultrahigh magnetic fields  {\it
  Journal of Magnetic Resonance. } 2020;313:106703.

\bibitem{perlman2020ai}
Perlman Or, Ito Hirotaka, Herz Kai, et al. AI boosted molecular MRI for
  apoptosis detection in oncolytic virotherapy  {\it bioRxiv. } 2020.

\bibitem{kim2009water}
Kim Mina, Gillen Joseph, Landman Bennett~A, Zhou Jinyuan, Van~Zijl Peter~CM.
  Water saturation shift referencing (WASSR) for chemical exchange saturation
  transfer (CEST) experiments  {\it Magnetic Resonance in Medicine: An Official
  Journal of the International Society for Magnetic Resonance in Medicine. }
  2009;61:1441--1450.

\bibitem{liu2010high}
Liu Guanshu, Gilad Assaf~A, Bulte Jeff~WM, Van~Zijl Peter~CM, McMahon
  Michael~T. High-throughput screening of chemical exchange saturation transfer
  MR contrast agents  {\it Contrast media \& molecular imaging. }
  2010;5:162--170.

\bibitem{zaiss2018quesp}
Zaiss Moritz, Angelovski Goran, Demetriou Eleni, McMahon Michael~T, Golay
  Xavier, Scheffler Klaus. QUESP and QUEST revisited--fast and accurate
  quantitative CEST experiments  {\it Magnetic resonance in medicine. }
  2018;79:1708--1721.

\bibitem{lein2007genome}
Lein Ed~S, Hawrylycz Michael~J, Ao~Nancy, et al. Genome-wide atlas of gene
  expression in the adult mouse brain  {\it Nature. } 2007;445:168--176.

\bibitem{AllenMouseIm78}
© 2004 Allen Institute for Brain Science. Allen Mouse Brain Atlas. Available
  from:
  \url{https://atlas.brain-map.org/atlas?atlas=1#atlas=1&plate=100960088}.

\bibitem{virtanen2020scipy}
Virtanen Pauli, Gommers Ralf, Oliphant Travis~E, et al. SciPy 1.0: fundamental
  algorithms for scientific computing in Python  {\it Nature methods. }
  2020;17:261--272.

\bibitem{van2003mechanism}
Van~Zijl Peter~CM, Zhou Jinyuan, Mori Noriko, Payen Jean-Francois, Wilson
  David, Mori Susumu. Mechanism of magnetization transfer during on-resonance
  water saturation. A new approach to detect mobile proteins, peptides, and
  lipids  {\it Magnetic Resonance in Medicine: An Official Journal of the
  International Society for Magnetic Resonance in Medicine. } 2003;49:440--449.

\bibitem{chen2018separating}
Chen Lin, Xu~Xiang, Zeng Haifeng, et al. Separating fast and slow exchange
  transfer and magnetization transfer using off-resonance variable-delay
  multiple-pulse (VDMP) MRI  {\it Magnetic resonance in medicine. }
  2018;80:1568--1576.

\bibitem{van2017rapid}
Gelderen Peter, Jiang Xu, Duyn Jeff~H. Rapid measurement of brain
  macromolecular proton fraction with transient saturation transfer MRI  {\it
  Magnetic resonance in medicine. } 2017;77:2174--2185.

\bibitem{herz2021pulseq}
Herz Kai, Mueller Sebastian, Perlman Or, et al. Pulseq-CEST: Towards multi-site
  multi-vendor compatibility and reproducibility of CEST experiments using an
  open-source sequence standard  {\it Magnetic resonance in medicine. } 2021.

\bibitem{stanisz2005t1}
Stanisz Greg~J, Odrobina Ewa~E, Pun Joseph, et al. T1, T2 relaxation and
  magnetization transfer in tissue at 3T  {\it Magnetic Resonance in Medicine:
  An Official Journal of the International Society for Magnetic Resonance in
  Medicine. } 2005;54:507--512.

\bibitem{liu2013quantitative}
Liu Dapeng, Zhou Jinyuan, Xue Rong, Zuo Zhentao, An~Jing, Wang Danny~JJ.
  Quantitative characterization of nuclear overhauser enhancement and amide
  proton transfer effects in the human brain at 7 tesla  {\it Magnetic
  resonance in medicine. } 2013;70:1070--1081.

\bibitem{kang2021unsupervised}
Kang Beomgu, Kim Byungjai, Sch{\"a}r Michael, Park HyunWook, Heo Hye-Young.
  Unsupervised learning for magnetization transfer contrast MR fingerprinting:
  Application to CEST and nuclear Overhauser enhancement imaging  {\it Magnetic
  resonance in medicine. } 2021.

\bibitem{kim2020deep}
Kim Byungjai, Sch{\"a}r Michael, Park HyunWook, Heo Hye-Young. A deep learning
  approach for magnetization transfer contrast MR fingerprinting and chemical
  exchange saturation transfer imaging  {\it NeuroImage. } 2020;221:117165.

\bibitem{zhu2018automated}
Zhu Bo, Liu JZ, Koonjoo Neha, Rosen B, Rosen M. AUTOmated pulse SEQuence
  generation (AUTOSEQ) using Bayesian reinforcement learning in an MRI physics
  simulation environment  in {\it Proc. 26th Annu. Meeting ISMRM}:16--21 2018.

\bibitem{loktyushin2021mrzero}
Loktyushin A, Herz K, Dang N, et al. MRzero-Automated discovery of MRI
  sequences using supervised learning  {\it Magnetic Resonance in Medicine. }
  2021;86:709--724.

\bibitem{glang2021advances}
Glang F, Loktyushin A, Herz K, et al. Advances in MRzero: supervised learning
  of parallel imaging sequences including joint non-Cartesian trajectory and
  flip angle optimization  in {\it 2021 ISMRM \& SMRT Annual Meeting \&
  Exhibition (ISMRM 2021)} 2021.

\bibitem{cohen2017algorithm}
Cohen Ouri, Rosen Matthew~S. Algorithm comparison for schedule optimization in
  MR fingerprinting  {\it Magnetic resonance imaging. } 2017;41:15--21.

\bibitem{wang2020novel}
Wang Zhenxiong, Shaghaghi Mehran, Zhang Shun, et al. Novel proton exchange rate
  MRI presents unique contrast in brains of ischemic stroke patients  {\it
  Journal of Neuroscience Methods. } 2020;346:108926.

\bibitem{meissner2015quantitative}
Meissner Jan-Eric, Goerke Steffen, Rerich Eugenia, et al. Quantitative pulsed
  CEST-MRI using $\Omega$-plots  {\it NMR in Biomedicine. } 2015;28:1196--1208.

\bibitem{roeloffs2015towards}
Roeloffs Volkert, Meyer Christian, Bachert Peter, Zaiss Moritz. Towards
  quantification of pulsed spinlock and CEST at clinical MR scanners: an
  analytical interleaved saturation--relaxation (ISAR) approach  {\it NMR in
  Biomedicine. } 2015;28:40--53.

\bibitem{gochberg2018towards}
Gochberg Daniel~F, Does Mark~D, Zu~Zhongliang, Lankford Christopher~L. Towards
  an analytic solution for pulsed CEST  {\it NMR in Biomedicine. }
  2018;31:e3903.

\bibitem{zaiss2011quantitative}
Zai{\ss} Moritz, Schmitt Benjamin, Bachert Peter. Quantitative separation of
  CEST effect from magnetization transfer and spillover effects by
  Lorentzian-line-fit analysis of z-spectra  {\it Journal of magnetic
  resonance. } 2011;211:149--155.

\bibitem{zhang2017accuracy}
Zhang Xiao-Yong, Wang Feng, Li~Hua, et al. Accuracy in the quantification of
  chemical exchange saturation transfer (CEST) and relayed nuclear Overhauser
  enhancement (rNOE) saturation transfer effects  {\it NMR in biomedicine. }
  2017;30:e3716.

\end{thebibliography}
\bibliographystyle{ama}

\newpage
\section*{Tables and figures}

\begin{table}[ht!]
\fontsize{11}{11}
\selectfont
  \begin{center}
    \caption{AutoCEST determined semi-solid and amide proton chemical exchange rates (k) and volume fractions (f) for GM and WM regions of in vivo mouse brain tissue.}
    \label{Table1}
    \begin{tabular}{p{4cm}  p{4cm}  p{4cm} }
    
      & \textbf{Cortical GM} & \textbf{Sub-cortical WM$^a$} 
       \\ \hline
Semi-solid k$_{ssw}$ (Hz) & 60.81$\pm$9.28 & 46.23$\pm$14.70\\ \hline
	  Semi-solid f$_{ss}$ ($\%$) & 12.21$\pm$1.37 & 19.73$\pm$3.30\\ \hline
	  Amide k$_{sw}$ (Hz) & 61.03$\pm$29.24 & 73.02$\pm$51.11\\ \hline
	 Amide f$_{s}$ ($\%$) & 0.29$\pm$0.16 & 0.40$\pm$0.27\\
    \end{tabular}
  \end{center}
\end{table}
{\small
\noindent Abbreviations: GM=gray matter, WM=white matter. \\
$^a$Average of corpus callosum and white matter fiber tracts (cerebal peduncle, optic tract, and fimbria) \cite{lein2007genome, AllenMouseIm78}.
}

\newpage
    
\newpage

\newpage

\newpage

\newpage

\newpage

\newpage

\newpage

\newpage

\end{document}